\documentclass[conference]{IEEEtran}

\IEEEoverridecommandlockouts

\usepackage{cite}
\usepackage{amsmath,amssymb,amsfonts}
\usepackage{graphicx}
\usepackage{textcomp}
\usepackage{xcolor}
\usepackage{comment}
\usepackage{hyperref}

\usepackage{grumble}
\usepackage{wrapfig}

\newcommand{\myparagraph}[1]{\noindent{\bf {#1}.}}
\newcommand*\circled[1]{\tikz[baseline=(char.base)]{
            \node[shape=circle,draw,inner sep=1pt] (char) {#1};}}

\newcommand{\takeaway}[1]{%
    \vspace{1mm}%
    \noindent\fbox{\parbox{\dimexpr\columnwidth-2\fboxsep-2\fboxrule\relax}{%
        \setlength{\parskip}{0pt}%
        \vspace{-2pt}%
        #1%
        \vspace{-2pt}%
    }}%
    \vspace{1mm}%
}

\newcommand{\projectname}{Chipmunq}

\usepackage[inline]{enumitem}
\setlist{noitemsep,topsep=0pt,parsep=0pt,partopsep=0pt}

\usepackage{setspace}
\usepackage[margin=5pt,font={stretch=0.9}]{caption}

\def\BibTeX{{\rm B\kern-.05em{\sc i\kern-.025em b}\kern-.08em‚
    T\kern-.1667em\lower.7ex\hbox{E}\kern-.125emX}}

\begin{document}
    
    \title{Chipmunq: A Fault-Tolerant Compiler for Chiplet Quantum Architectures}

    \author{
        \IEEEauthorblockN{Peter Wegmann, Aleksandra Świerkowska, Emmanouil Giortamis, and Pramod Bhatotia}
        \IEEEauthorblockA{Technical University of Munich, Munich, Germany \\
        \{peter.wegmann, aleksandra.swierkowska\}@tum.de}
    }
    
    \maketitle
    
    \begin{abstract}
        As quantum computing advances toward fault-tolerance through quantum error correction, modular chiplet architectures have emerged to provide the massive qubit counts required while overcoming fabrication limits of monolithic chips. However, this transition introduces a critical compilation gap: existing frameworks cannot handle the scale of fault-tolerant quantum circuits while managing the noisy, sparse interconnects of chiplet backends. 

We present \projectname{}, the first hardware-aware compiler for mapping and routing fault-tolerant circuits onto modular architectures. \projectname{} employs a quantum-error-correction-aware partitioning strategy that preserves the integrity of logical qubit patches, preventing prohibitive gate overheads common in general-purpose compilers. Our evaluation demonstrates that \projectname{} achieves a 13.5$\times$ speedup in compilation time compared to state-of-the-art tools. By incorporating chiplet constraints and defective qubits, it reduces circuit depth by 86.4\% and SWAP gate counts by 91.4\% across varying code distances. Crucially, \projectname{} overcomes heterogeneous inter-chiplet links, improving logical error rates by up to two orders of magnitude.
    \end{abstract}
    
    
    \section{Introduction}
    \label{sec:introduction}
    
    \myparagraph{Context and motivation}
    Quantum computing currently operates in the Noisy Intermediate-Scale Quantum (NISQ) era \cite{Preskill2018}, characterized by two fundamental bottlenecks: \textit{noise} arising from fabrication defects and hardware imperfections \cite{devoret2004superconducting, hertzberg2021laser}, and \textit{limited scale} constrained by the exponential growth in control complexity and cryogenic cooling requirements as qubit counts increase \cite{tian2025youtiao, fruitwala2024distributed, smith2022scaling}. 

    To address noise and enable fault-tolerant quantum computing (FTQC), {\em Quantum Error Correction (QEC)} has emerged as the cornerstone strategy \cite{aharonov2008fault, Fowler_2012, shor1996fault}. QEC encodes logical quantum information redundantly across multiple physical qubits, enabling error detection and correction through syndrome measurements without collapsing the quantum state \cite{gottesman1998theory, gottesman2010introduction}. 
    However, this reliability comes at a steep cost: QEC dramatically amplifies resource demands, increasing both qubit counts and circuit depth by orders of magnitude as code distance scales to suppress logical error rates \cite{Fowler_2012}.
    
\begin{figure*}[t!]
    \centering
    \includegraphics[width=\linewidth]{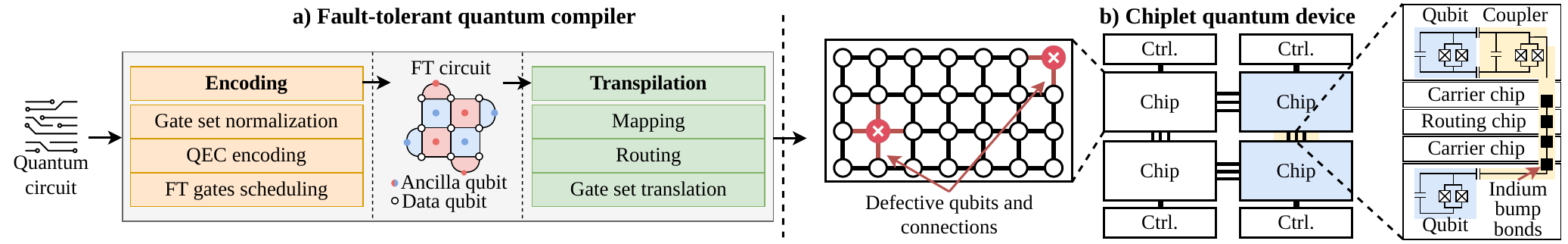}
    
    \caption{Example of the compilation process in which a quantum circuit is converted into FT form targeting a quantum chiplet architecture.
    \label{fig:background_combined}}
    \vspace{-15pt}

\end{figure*}

    To address scale and bypass the integration limits of monolithic processors, quantum hardware is rapidly transitioning toward {\em chiplet-based architectures} \cite{smith2022scaling, wu2024modular, laracuente2025modeling}. Rather than fabricating all qubits on a single die, chiplet designs interconnect multiple smaller quantum processing units (QPUs) via dedicated inter-chip links such as microwave couplers or 3D bump bonds (Fig. \ref{fig:background_combined} (b)). This modular approach improves fabrication yield, simplifies control wiring, and enables incremental scaling \cite{smith2022scaling}. Yet it introduces a new challenge: inter-chiplet links are typically sparse, slower, and exhibit noise rates 10-100$\times$ higher than intra-chip couplings, creating significant architectural heterogeneity \cite{ibm-remote-gate-cost, norris2026performance}.
    
    \myparagraph{Research challenges} The convergence of QEC and chiplet architectures, while necessary for practical quantum computing, creates a ``compilation wall'' defined by three interrelated challenges, which we quantitatively motivate in \S~\ref{sec:motivation}:

    \noindent
    {\em (1) Exponential resource explosion (\S~\ref{sec:motivation:scalability}):}
    Existing qubit-level compilation heuristics, designed for small, monolithic devices, become computationally intractable at the scale of fault-tolerant (FT) circuits and chiplet QPUs, often timing out or requiring hours for code distances approaching practical thresholds $d \approx 27$~\cite{Acharya2025}.

    \noindent
    {\em (2) QEC patch fragmentation (\S~\ref{sec:motivation:patch-preserving}):} Surface codes organize physical qubits into discrete geometric patches, each representing a single logical qubit \cite{Fowler_2012}. Multi-qubit logical operations, such as lattice surgery CNOTs, require carefully coordinated interactions between spatially localized patches \cite{cnot_lattice_surgery}. Standard compilers optimize gate proximity at the individual qubit level without awareness of this higher-level structure. As a result, they frequently fragment patches across noisy chiplet boundaries or introduce excessive SWAP operations that disrupt patch geometry, directly undermining the error-suppression capabilities that QEC is designed to provide \cite{eccentric}.

    \noindent
    {\em (3) Interconnect heterogeneity (\S~\ref{sec:motivation:partitioning}):}
    General-purpose mapping and routing compilers treat all qubit links uniformly, often routing critical QEC syndrome measurements over low-fidelity inter-chiplet links. This mismatch between circuit requirements and hardware reality can render QEC ineffective, leading to logical error rates exceeding correction thresholds even when the underlying QEC code would otherwise succeed.

    \myparagraph{Research gap} Despite recent advances in quantum compilation, no existing framework simultaneously addresses scalability, patch preservation, and chiplet-aware placement. Chiplet-dedicated compilers like MECH \cite{mech} and general-purpose tools like LightSABRE \cite{ibm_sabre} operate at the physical qubit level, failing to scale efficiently or preserve QEC structure. QEC-aware compilers such as QECC-Synth \cite{qecc_synth} maintain patch integrity but time out beyond modest code distances and do not account for inter-chiplet noise heterogeneity. This leaves a critical void: the lack of a compiler capable of mapping FT quantum circuits onto modular chiplet architectures in a scalable, structure-preserving, and architecture-aware manner.

    This work addresses the following central question: \textit{How can fault-tolerant quantum circuits be mapped and routed onto chiplet-based architectures in a scalable, QEC patch-preserving, and architecture-aware manner?}

    \myparagraph{Our key ideas}
    Realizing such a compiler requires rethinking compilation at three levels: 
    
    \noindent
    {\em (1) QEC-aware circuit decomposition:} Rather than partitioning circuits naively by qubit or chiplet count, which fragments patches and increases inter-chiplet communication, we apply community detection to the circuit interaction graph to identify QEC patches first. Partitioning decisions are then made exclusively at inter-patch boundaries. This patch-centric abstraction reduces the effective problem size from thousands of qubits to tens of patches, enabling the use of high-quality partitioning solvers that would otherwise be intractable at the physical qubit level \cite{dqc_np_hard}.
    
    \noindent
    {\em (2) Architecture-aware partition placement:} Assigning partitions to physical chiplets is a complex bin-packing problem under capacity and connectivity constraints \cite{lattice_surgery_np_hard}. We decompose this into two coordinated stages: a sequencer constructs a dependency-aware execution order that prioritizes highly connected partitions, and a global mapper performs localized, defect-aware 2D bin-packing assignments that respect this ordering. By staging the optimization, we avoid the exponential search space of joint optimization while preventing the placement density issues common in greedy heuristics.
    
    \noindent
    {\em (3) Connectivity-constrained circuit realization:} Enforcing nearest-neighbor connectivity at scale requires SWAP insertion that respects both intra-chiplet patch geometry and inter-chiplet link quality. We implement a hierarchical, cost-aware router that preserves patch shapes during local mapping, then applies a weighted cost function during inter-chiplet routing that balances path length, link fidelity, and congestion. This dual-stage approach limits cascading inefficiencies and ensures that critical syndrome measurements avoid low-fidelity communication paths.

    \myparagraph{Our approach} We present \projectname{}, the first hardware-aware compilation framework designed to map and route FT quantum circuits onto modular chiplet architectures. \projectname{} is structured as a modular five-stage pipeline: (1) a \textit{partitioner} using community detection to identify and preserve QEC patches (\S~\ref{subsec:implementation_partitioning}); (2) a \textit{sequencer} to establish dependency-aware execution ordering (\S~\ref{subsec:implementation_mapping}); (3-4) an architecture-aware \textit{mapper} (global and local) utilizing 2D bin-packing with defect awareness (\S~\ref{subsec:implementation_mapping}); and (5) a cost-aware \textit{router} that explicitly models inter-chiplet link fidelity and congestion (\S~\ref{subsec:implementation_routing}).  To facilitate seamless information flow across these compilation stages, we introduce the \projectname{} Intermediate Representation (IR) (\S~\ref{subsec:ir}), which couples an immutable logical dependency graph with a progressively enriched partition registry. As the circuit traverses the pipeline, each stage appends hardware-binding metadata without modifying the underlying logical structure.

    \myparagraph{Implementation and evaluation} We implement \projectname{} as a custom pass manager pipeline integrated into Qiskit v2.30rc1 \cite{qiskit}, leveraging KaHyPar \cite{kahypar} for partitioning  and rustworkx for graph algorithms \cite{rustworkx}. We evaluate \projectname{} using surface-code-protected lattice surgery benchmarks generated via tQEC \cite{tqec}, simulating compiled circuits with Stim \cite{stim_gidney} to measure logical error rates (LER). Our evaluation spans circuits of varying complexity (1-8 CNOTs) and code distances (d=3-15), targeting diverse chiplet backends with varying interconnect densities and heterogeneous noise profiles. \projectname{} achieves average speedup of 13.5$\times$ in compilation time compared to state-of-the-art general-purpose compilers like LightSABRE \cite{ibm_sabre}, while reducing circuit depth by 86.4\% and SWAP gate counts by 91.4\% on average. Crucially, \projectname{}'s noise-aware routing suppresses logical error rates by up to two orders of magnitude in heterogeneous environments, demonstrating that chiplet modularity need not come at the cost of fault-tolerance fidelity.
    
    \myparagraph{Contributions} Our contributions are as follows:
    
    \begin{itemize}
        \item {\em The \projectname{} compiler:} The first scalable, hardware-aware mapping and routing framework that bridges the gap between topological QEC codes and heterogeneous chiplet-based architectures.
    
        \item {\em Noise-aware heuristics:}  A novel inter-chiplet routing strategy that incorporates link fidelity and congestion awareness, successfully suppressing logical error rates by up to two orders of magnitude in heterogeneous noise environments while maintaining QEC effectiveness below correction thresholds.   
        
        \item {\em Scalability benchmarking:} An evaluation demonstrating that \projectname{} delivers 13.5$\times$ speedup on average in compilation time and an average 91.4\% reduction in SWAP overhead compared to state-of-the-art baselines.
            
    \end{itemize}

    \begin{figure*}[t!]
    \centering
    \includegraphics[width=.3\linewidth]{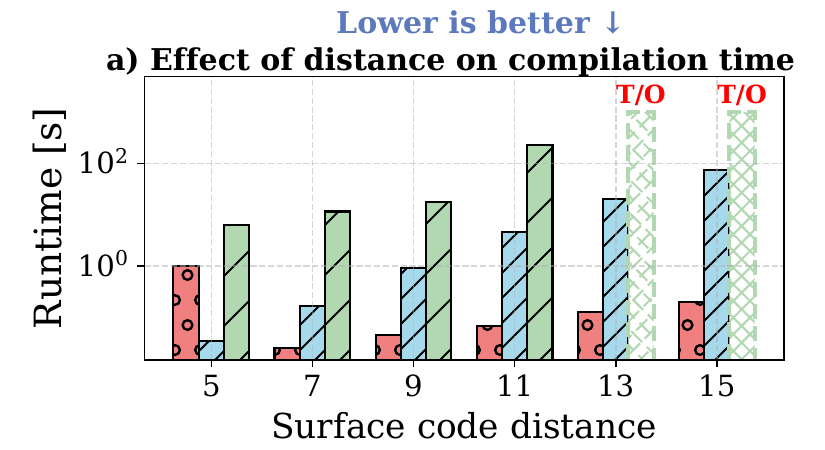}
    \includegraphics[width=.3\linewidth]{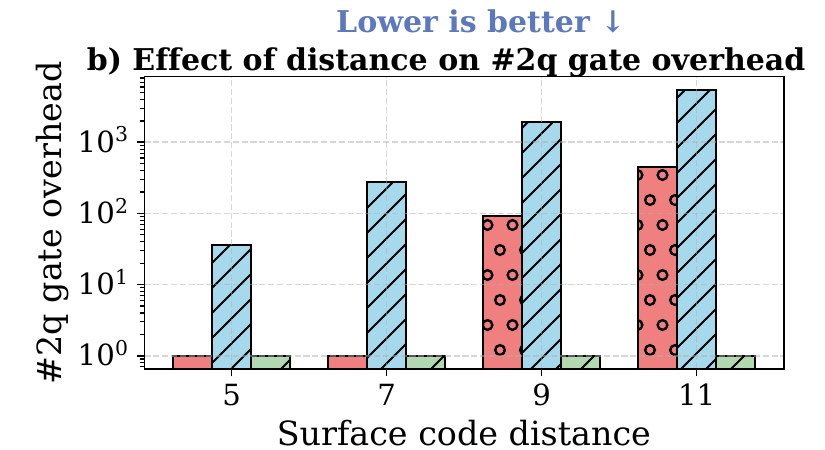}
    \includegraphics[width=.3\linewidth]{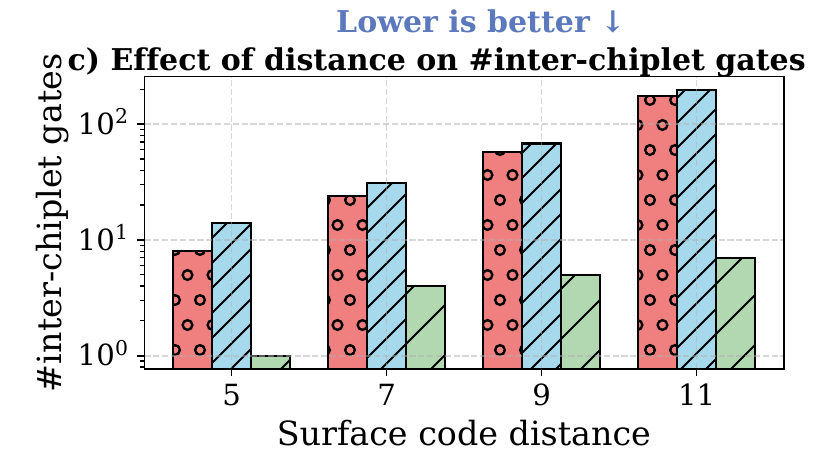}
    
    \centering
    \includegraphics[width=.33\linewidth]{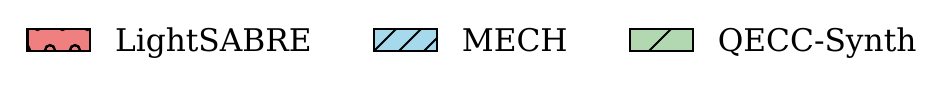}

    \vspace{-5pt}
    \caption{Comparison of compiling a single surface code patch by LightSABRE~\cite{ibm_sabre}, MECH~\cite{mech}, and QECC-Synth~\cite{qecc_synth} compilers. \emph{
    \textbf{(a)} Compilation runtime with runs exceeding $10^3$~s reported as timeouts (T/O). 
    \textbf{(b)} Two-qubit gate overhead introduced when compiling to a monolithic backend. 
    \textbf{(c)} Number of two-qubit gates executed over inter-chiplet connections when compiling to a chiplet backend.}} 
    \label{fig:motivation_comparison}
    \vspace{-7pt}
    
\end{figure*}

\section{Background}\label{sec:background}
    In this section, we present the technical foundations of this work. We first examine quantum chiplet architectures, a promising approach to addressing the scalability constraints of NISQ devices. Next, we discuss FTQC and its role in addressing the noise challenge of NISQ devices. Last, we discuss the quantum compilation pipeline in the unique context of chiplet constraints and FTQC requirements.
    
    \subsection{Quantum Chiplet Architectures}\label{sec:background:chiplets}
        Quantum computing platforms utilize diverse technologies \cite{bluvstein2024logical, haffner2008quantum, psiquantum2025manufacturable}, yet superconducting qubits remain the state-of-the-art \cite{arute2019quantum} due to rapid gate operations and compatibility with standard lithographic fabrication \cite{superconducting_transmon}. To overcome the scalability bottlenecks of monolithic processors — declining fabrication yields \cite{monolithic_to_chiplets}, increasing control complexity \cite{fruitwala2024distributed}, and wiring constraints \cite{tian2025youtiao} — recent hardware designs have adopted \emph{chiplet-based} architectures \cite{monolithic_to_chiplets, kurpiers2018deterministic, zhong2021deterministic, magnard2020microwave, gold2021entanglement, smith2022scaling, wu2024modular, laracuente2025modeling}. In a chiplet architecture, multiple smaller quantum chips are interconnected using dedicated inter-chiplet links over which quantum gates or state transfer operations can be performed \cite{chiplet_interconnects}, as illustrated in Fig.~\ref{fig:background_combined} (b). However, these benefits introduce new challenges. In particular, inter-chiplet communication is typically noisier, highly constrained, and less uniform than intra-chip connectivity, leading to architectural heterogeneity that must be explicitly accounted for at higher levels of the system stack \cite{ibm-remote-gate-cost, norris2026performance, monolithic_to_chiplet_links}. 
        
    \subsection{Fault-Tolerant Quantum Computing (FTQC)}\label{sec:background:ftqc}
        To mitigate the noise inherent in NISQ-era devices, QEC is essential for achieving FTQC \cite{Fowler_2012}. QEC protects quantum information by encoding a single logical qubit into an entangled state of multiple physical qubits, enabling error detection and correction via parity measurements without collapsing the underlying quantum state \cite{terhal2015quantum}. There exist various QEC codes with distinct trade-offs between error thresholds, resource overhead, and hardware compatibility \cite{breuckmann2021quantum, bombin2006topological}, with a notable example being the surface code \cite{Acharya2025}, which is widely adopted due to its high error threshold and reliance on 2D local connectivity.
        
        In all QEC codes, the physical qubits are typically arranged in discrete patches, which serve as the fundamental units for logical operations, as shown in Fig.~\ref{fig:background_combined} (a). This patch-based representation fundamentally changes how quantum operations are implemented. While single-logical-qubit operations can often be performed locally within a patch, multi-logical-qubit gates require carefully coordinated interactions between multiple patches. In the surface code, these interactions are typically implemented via lattice-surgery operations, such as merges and splits \cite{Fowler_2012}. This patch-based structure fundamentally changes compilation requirements: compilers must preserve patch geometry while managing spatial layout, temporal scheduling, and resource constraints, significantly increasing the need for specialized compilation techniques for FTQC programs.
        
    \subsection{Quantum Compilation}\label{sec:background:compilation}
        Quantum compilation is the process of translating architecture-independent quantum algorithms into hardware-specific instructions that satisfy the physical and topological constraints of the target backend. The standard workflow involves \textit{synthesis} to decompose high-level gates, \textit{mapping} to assign logical qubits to physical locations \cite{swamit2019not, tannu2019ensemble}, and \textit{routing} to satisfy connectivity constraints \cite{gushu2019tackling, cowtan2019qubit}, the latter two being NP-hard problems \cite{SiraichiSCP18, cowtan2019qubit}, concluding with \textit{gate-set translation} into native hardware operations \cite{yunong2019optimized, das2023the}.
        
        For FT operations, as shown in Fig.~\ref{fig:background_combined} (a), the pipeline begins with an encoding phase that includes \textit{gate set normalization}, \textit{QEC encoding}, and \textit{FT gate scheduling}. This phase generates an intermediate FT circuit composed of physical qubit patches. This circuit then undergoes transpilation, as stated before, to be optimized for the target architecture. The integration of QEC-encoded patches with modular chiplet architectures introduces unique scaling and routing challenges due to the significantly increased complexity of the interaction graphs, which we address in the following sections.

    \begin{figure*}[t]
    \centering
    \includegraphics[width=0.9\linewidth]{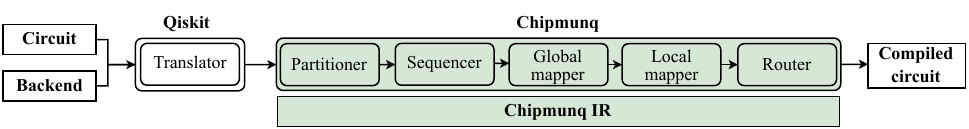}
    \caption{Overview of \projectname{}. \emph{Our modular approach integrates with Qiskit and supports circuit-level and QEC-specific simulations.}}
    \label{fig:architecture}
    \vspace{-5pt}
\end{figure*}

\section{Motivation}\label{sec:motivation}
    Recent advances in chiplet quantum architectures and QEC are bringing quantum computing closer to the FTQC era. 
    However, the combination of large patch-based FT circuits and noisy, limited inter-chiplet connectivity stresses highly complex, qubit-level compilation, increasing its runtime and overhead. 
    In this section, we empirically evaluate the impact of chiplet characteristics and FT circuit structure on mapping and routing, and highlight how state-of-the-art compilers handle these challenges.
            
    \subsection{Low Scalability} \label{sec:motivation:scalability}
        
        Since qubit routing is an NP-hard problem~\cite{SiraichiSCP18}, scaling compilation in the context of FT circuits and chiplet architectures requires highly efficient heuristics that balance a delicate trade-off between solution quality and practical runtime. 
        
        

        In Fig.~\ref{fig:motivation_comparison} (a), we evaluate the scalability of LightSABRE~\cite{ibm_sabre}, MECH~\cite{mech}, and QECC-Synth~\cite{qecc_synth} using surface-code memory circuits~\cite{Fowler_2012} for increasing code distances. While QECC-Synth times out at d=11 and MECH shows linear runtime growth, LightSABRE maintains sub-second execution across all evaluated sizes. Since practical FTQC necessitates $d \approx 27$~\cite{Acharya2025}, these results highlight the scalability limitations of current QEC- and chiplet-specific compilers for large-scale workloads.
        
        \takeaway{\textbf{Takeaway:} QEC- and chiplet-specific algorithms struggle to scale to large FT circuits within acceptable compilation time, leaving only general-purpose compilers practical at larger code distances.}

    \subsection{Patch-Preserving Mapping and Routing}\label{sec:motivation:patch-preserving}
        
        Mapping and routing for FT execution must preserve the structural constraints imposed by QEC codes to maintain their error-suppression properties, i.e., they should not alter patch geometry or introduce additional SWAP operations that degrade FT reliability.
        
        We quantify routing overhead in Fig.~\ref{fig:motivation_comparison} (b) by reporting the additional two-qubit gates introduced when mapping QEC patches onto a monolithic grid. Since the target architecture matches the patch geometry, an ideal compiler should preserve patch structure with zero overhead. We observe that QECC-Synth alone maintains this structure across all distances. In contrast, LightSABRE begins violating patch integrity at $d=9$, leading to rapidly increasing overhead, while MECH introduces nearly linear gate overhead across all evaluated distances.
        
        \takeaway{\textbf{Takeaway:} Mapping and routing approaches that are not explicitly QEC-aware fail to preserve patch structure, introducing significant two-qubit gate overhead that degrades FT reliability.}
    
    \subsection{Architecture-Aware Partitioning}\label{sec:motivation:partitioning}
        
        FT circuits must be partitioned across chiplets in a manner that minimizes costly inter-chip communication and does not directly or indirectly increase error rates through additional SWAP insertion.

        In Fig.~\ref{fig:motivation_comparison} (c), we evaluate how different compilers perform mapping and routing of QEC code patches on a chiplet-based architecture by measuring the number of two-qubit gates executed over inter-chiplet connections. To isolate partitioning behavior, we scale the chiplet size so that each patch fits entirely within a single chiplet. Despite this, both the general-purpose LightSABRE and the chiplet-dedicated MECH exhibit approximately linear growth in inter-chiplet two-qubit operations as circuit size increases, indicating heavy reliance on these noisy links. Although QECC-Synth achieves lower inter-chiplet utilization by exploiting mechanisms designed for sparse architectures, it still introduces a non-negligible and growing number of inter-chiplet operations.
        
        \takeaway{\textbf{Takeaway:} All compilers unnecessarily utilize noisy inter-chiplet connections for FT workloads, highlighting the insufficiency of existing partitioning strategies.}
        
    \subsection{Problem Statement}
        \noindent\fbox{\parbox{\dimexpr\columnwidth-2\fboxsep-2\fboxrule\relax}{%
                \setlength{\parskip}{0pt}%
                \vspace{-2pt}%
                \emph{How can FT quantum circuits be mapped and routed onto chiplet-based architectures in a scalable, QEC patch-preserving, and architecture-aware manner?}%
                \vspace{-2pt}%
            }}%

    \section{Chipmunq Overview} \label{sec:overview}
    
\begin{figure*}[t]
    \centering
    \includegraphics[width=0.9\linewidth]{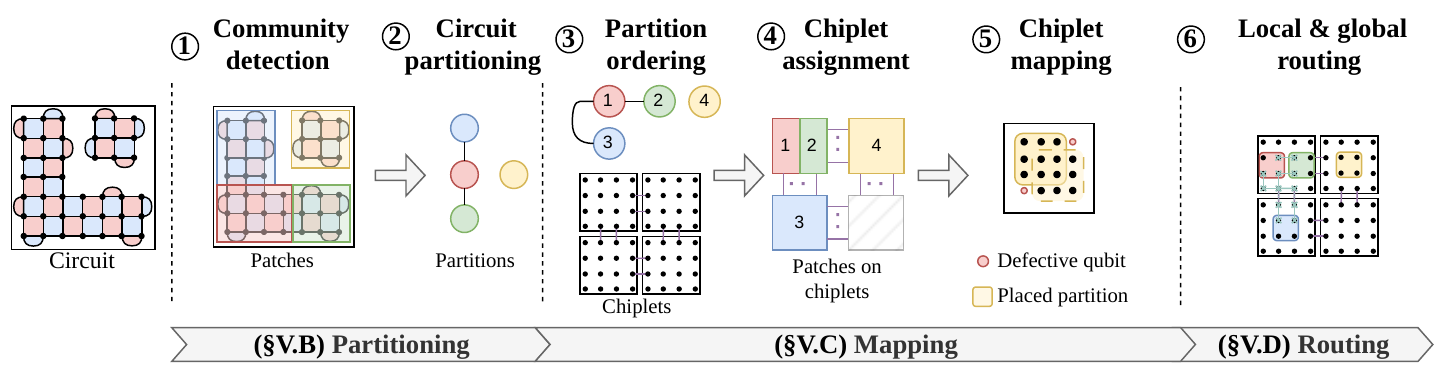}
    \caption{\projectname's workflow given a step-by-step example. \emph{An example input circuit is compiled to a chiplet backend, where each chiplet is assigned one or multiple patches.}}
    \label{fig:workflow}
    \vspace{-7pt}
\end{figure*}
    
    In this work, we propose \projectname{}, a compiler that enables mapping and routing FT circuits onto both monolithic and chiplet-based architectures in a scalable, QEC-patch-preserving, and architecture-aware manner, aligning with the evolving demands of near- and long-term quantum modular hardware architectures.      
      
    To achieve this, we design our compiler with four primary design goals: 
    \circled{1} \textbf{Error suppression:} \projectname{} prioritizes QEC effectiveness by minimizing unnecessary noisy SWAP operations and limiting the use of lower-fidelity links in heterogeneous hardware setups. This architecture-aware approach improves the reliability of FT computation.
    \circled{2} \textbf{Scalability:} The compiler is capable of handling the increasing size and complexity of FT circuits as well as large-scale, post-NISQ quantum processors, ensuring practical deployment across diverse hardware platforms.
    \circled{3} \textbf{Generality:} \projectname{} supports arbitrary QEC codes and can adapt to various chiplet sizes and layouts, making it applicable to a wide range of hardware architectures.
    \circled{4} \textbf{Modularity:} Our compiler cleanly separates high-level FT compilation (e.g., lattice surgery) from low-level mapping and routing to backends. Thus, \projectname{} eliminates the need to re-run high-level compilation across hardware targets and enables concurrent compilation of multiple FT circuits to a single backend.

    \subsection{Chipmunq Architecture}\label{sec:overview:architecture}
        \projectname{} takes as input a high-level FT circuit in Qiskit \texttt{QuantumCircuit} format \cite{qiskit} and returns a fully compiled circuit as a Directed Acyclic Graph (DAG) encoding all mapping and routing decisions for hardware execution. To achieve this, \projectname{} performs the five steps presented in Fig.~\ref{fig:architecture}:

        Firstly, to minimize high-latency inter-chiplet communication, the \textbf{partitioner} clusters circuit operations by analyzing the circuit’s DAG and applying community detection algorithms to group operations into partitions that align with QEC patches, thereby preserving internal locality. 
       
        Then, the \textbf{sequencer} establishes a mapping order that maximizes placement flexibility for the most interconnected partitions. By building a dependency graph over the defined partitions, it derives a global linear ordering that prioritizes highly connected components to reduce downstream routing complexity.
        
        Once we fix the order, we assign partitions to distinct chiplets to minimize inter-chiplet communication. The \textbf{global mapper} arranges the partitions based on their size, dependencies, chiplet capacity, and inter-chiplet connectivity.
       
        Subsequently, to avoid hardware defects and minimize internal data movement, the \textbf{local mapper} performs intra-chiplet placement. It leverages the known structure of QEC patches to map logical qubits onto physical qubits, effectively reducing the necessary SWAP operations while bypassing faulty hardware components.
        
        Finally, we enforce strict nearest-neighbor connectivity. As such, the \textbf{router} applies a heuristic to insert the SWAP operations needed to bring interacting qubits physically together. The router explicitly manages communication across chiplet interconnects in an error-aware manner and generates the final output DAG.
        
        To preserve information across these compilation stages, we introduce the \textbf{\projectname{} IR}. This IR couples the circuit DAG with a dynamic registry of hardware-bound partitions, enabling the system to propagate partition boundaries and logical ordering to subsequent passes while enriching the DAG with hardware-specific metadata.

    \subsection{Chipmunq Workflow}\label{sec:overview:workflow}
        We present \projectname{}'s workflow in Fig.~\ref{fig:workflow}. The compiler takes as input an FT quantum circuit with three logical qubits and a CNOT gate applied to two of them, utilizing an additional patch between the interacting qubits \cite{cnot_lattice_surgery}.
        
        We first run the partitioner, which analyzes the circuit to detect communities corresponding to QEC patches \textbf{(1)}, highlighted as differently colored regions. Based on this analysis, we divide the circuit into a fixed number of partitions while preserving information about their interdependencies \textbf{(2)}. The sequencer then uses this metadata to construct a dependency graph and derives an execution order that guides the subsequent backend mapping \textbf{(3)}.
        
        
        Given a target device with four chiplets, we next perform global mapping. We assign partitions according to their sizes and interconnectivity \textbf{(4)}. In this example, we place the most interconnected partition (Partition~1) together with the partition it communicates with most frequently (Partition~2) on the same chiplet to reduce communication overhead. We then apply local mapping to determine how to place the logical qubits of each partition onto physical qubits within the selected chiplet \textbf{(5)}. When mapping Partition~4, we avoid defective qubits at the chiplet's corners and preserve a square-like layout for the patch to maintain its structural properties.
        
        Finally, we invoke the router \textbf{(6)} to insert additional SWAP operations that enable communication both within chiplets (between Partitions~1 and~2) and across noisy inter-chiplet links (between Partitions~1 and~3). At the end, we produce a fully prepared circuit in DAG form that satisfies all hardware constraints and enables direct execution on the target backend.
    
\begin{figure*}[t]
    \centering
    \includegraphics[width=0.7\linewidth]{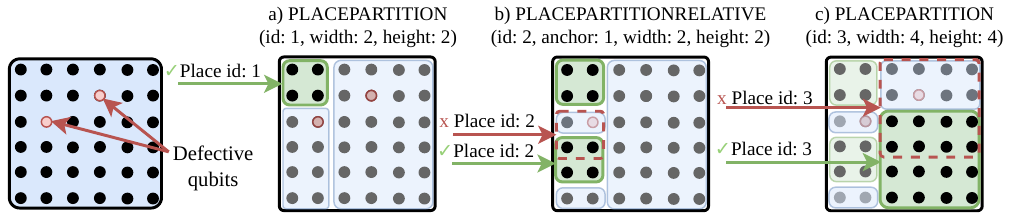}
    \caption{Example usage of \textsc{PlacePartition} and \textsc{PlacePartitionRelative} used in Algorithm \ref{alg:implementation_mapping} for placing differently sized partitions on a single chiplet containing defective qubits. \emph{\textbf{(a)} \textit{Size-aware} variant for placement of the first partition. \textbf{(b)} Placement of partitions that interact with each other. \textbf{(c)} Placement of partitions of different sizes in the presence of defective qubits. Green zones indicate successfully allocated partitions, red outline presents unsuccessful placements, and blue zones represent available regions.}}
    \label{fig:chiplet_mapping}
\end{figure*}
    
    \section{Chipmunq Design}
    
    We structure \projectname{} as a five-stage pipeline that optimizes FT circuits for chiplet architectures. We first introduce a hardware-aware IR (\S~\ref{subsec:ir}), followed by a partitioning stage that groups logical qubits into chiplet-scale domains (\S~\ref{subsec:implementation_partitioning}). We then detail a three-phase mapping stage (\S~\ref{subsec:implementation_mapping}) comprising the sequencer, global mapper, and local mapper. Finally, we present a communication-aware routing procedure (\S~\ref{subsec:implementation_routing}) that explicitly models inter-chiplet interconnects.
    
    \subsection{The Chipmunq Intermediate Representation}\label{subsec:ir}
        At the input boundary, the program is expressed as a Qiskit circuit with no spatial embedding, no notion of physical qubits, and no awareness of hardware defects or inter-chiplet communication limits \cite{qiskit}. At the output boundary, the program must be executable on a backend with fixed topology, bounded inter-chip bandwidth, heterogeneous reliability, and potential fabrication defects. Bridging this abstraction gap requires an IR that simultaneously preserves logical correctness while progressively incorporating hardware constraints.
    
        \myparagraph{Backend}
            We consider a chiplet architecture $C$ as a set $\{c_1, c_2, \dots, c_N\}$, with $N = 2^k$ and $k\in \mathbb{Z}$ as system size. We define $c_i$ as an individual monolithic chiplet connected to its neighbors in a rectangular grid. We define $n_{\text{icc}}$ number of inter-chiplet links $ICC$ as a list of tuples $(q_{icc, i}, \epsilon)$, with $q_{icc, i}$ as the local qubit $q_i$ supporting at most one inter-chiplet link to a neighboring chiplet with error rate $\epsilon$. A qubit $q_i$ can exist in one of two states: \textit{functional} or \textit{defective}. If $q_i$ is in the defective state, then the set of all incident connections is similarly mapped to the defective state.
        
       \myparagraph{Chipmunq IR}
            To mediate between the abstract circuit and the backend model, we introduce the \emph{Chipmunq IR}, an accumulative IR that couples an immutable logical dependency graph with a progressively enriched hardware-binding registry.
              
            The Chipmunq IR, $\mathcal{I}$, is defined as a 2-tuple $(G,\Pi)$. The \textbf{logical basis ($G$)} represents a Directed Acyclic Graph $G=(V,E)$, where $V$ is the set of quantum operations and $E$ represents qubit dependencies. The \textbf{partition registry ($\Pi$)} represents a set of stateful partition objects $\{\pi_1,\pi_2,\dots,\pi_n\}$, each of which is progressively enriched with hardware-binding metadata as it traverses the compiler pipeline.
            Each partition $\pi_i \in \Pi$ is defined as:
            \[
            \pi_i = \langle Q_i, \sigma_i, \alpha_i, \phi_i \rangle,
            \]
            where $Q_i \subset Q_{\mathrm{virt}}$ is the set of virtual qubits assigned to the partition.  $\sigma_i \in \mathbb{Z}_{\geq 0}$ is a scheduling annotation assigned during sequencing, encoding the partition's position in the global execution order relative to inter-partition dependencies. $\alpha_i$ is the chiplet assignment produced by the \textit{global mapper}, identifying which chiplet the partition is allocated to. $\phi_i : Q_i \rightarrow P(x,y)$ is the qubit coordinate mapping produced by the \textit{local mapper}, assigning each virtual qubit in $Q_i$ to a specific physical location within chiplet $\alpha_i$
        
            The compilation process is defined as a series of enrichment functions 
            $f : \mathcal{I}_k \rightarrow \mathcal{I}_{k+1}$ that strictly increase 
            the information density of $\Pi$, following the unidirectional trajectory:
            
            \[
            G \xrightarrow{f_{\mathrm{part}}} \Pi_{\{Q\}} 
              \xrightarrow{f_{\mathrm{seq}}} \Pi_{\{Q,\sigma\}} 
              \xrightarrow{f_{\mathrm{gmap}}} \Pi_{\{Q,\sigma,\alpha\}}
            \]
            \[
            \Pi_{\{Q,\sigma,\alpha\}} \xrightarrow{f_{\mathrm{lmap}}} \Pi_{\{Q,\sigma,\alpha,\phi\}} 
              \xrightarrow{f_{\mathrm{route}}} G_{\mathrm{final}}
            \]
            
            \noindent where partitioning decomposes $G$ into logical clusters with qubit assignments $Q_i$, sequencing resolves inter-partition dependencies to produce a global execution order $\sigma_i$, global mapping assigns each partition to a physical chiplet $\alpha_i$, local mapping places each virtual qubit at a coordinate within that chiplet $\phi_i$, and routing synthesizes $G_{\mathrm{final}}$ by inserting movements that satisfy the HW coupling map.

    \subsection{Partitioning}\label{subsec:implementation_partitioning}
        Utilizing chiplet architectures requires distributing the circuit across multiple chiplets. Our approach generates a circuit partitioning utilized by the subsequent mapping stage.
        To distribute the circuit across chiplets, we partition it into fixed groups. To preserve patches inherent to the circuit, we adopt a partition-aware approach that decomposes the circuit based on the number of QEC patches rather than the number of available physical chiplets. This strategy decouples the circuit partitioning from the backend. To generate an optimal partitioning, we employ a two-stage approach involving a community detection and a partitioning algorithm.

            To identify strongly connected sub-circuits, we apply the Girvan-Newman algorithm, an $\mathcal{O}(n^3)$ \textbf{community detection algorithm}, to a qubit entanglement graph constructed from the circuit \cite{qvm}, yielding a list of communities representing sub-circuits \cite{girvan_newman_community}. This approach is widely used in circuit cutting and tensor networks for partitioning circuits into smaller, more manageable subcircuits.
            Rather than using the raw partitions directly, we extract the total count of detected sub-circuits to estimate the target number of partitions for the subsequent steps. By decoupling the partition count calculation from the subsequent partitioning process, we can utilize algorithms that both require a fixed partition count and those that determine it dynamically.

            To construct the \textbf{circuit partitioning}, we utilize the estimated community count and size. We model this as a $k$-way partitioning problem, aiming to minimize the number of cut edges while adhering to fixed partition capacities. By employing the KaHyPar framework with near-logarithmic runtime on the constructed qubit entanglement graph, we minimize the connectivity metric, yielding an optimal circuit partitioning \cite{kahypar}.
            Since, for circuits prepared by high-level FT compilers (e.g., using surface code lattice surgery), we expect the patch assignments to be known beforehand, our framework supports predefined partitions via a dictionary that maps each circuit qubit to its partition ID. This mapping can then be used in the subsequent mapping step. 

    \subsection{Sequencing and Mapping}\label{subsec:implementation_mapping}
        Following the partitioning stage, we perform the mapping procedure, resulting in an assignment from virtual circuit qubits to physical qubits of the specified backend. We employ a three-step approach comprising the calculation of \textit{(1) sequencer}, \textit{(2) global mapper}, and \textit{(3) local mapper}, as shown in Algorithm \ref{alg:implementation_mapping}. 

\begin{algorithm}[t]
    \caption{Mapping and placement of partitions on chiplets}
    \label{alg:implementation_mapping}

    \small
    \begin{algorithmic}
        \Procedure{Mapping}{C, P}
            \State $P_{\text{ordered}}$ $\gets$ \textsc{PartitionSequencing}(P)
            \State $\pi_{C} \gets$ \textsc{GlobalMapping}(C, $P_{\text{ordered}}$)
            \State QubitMapping $\gets$ \textsc{ChipletMapping}(C, $\pi_{C}$)
            
            \State \textbf{return} QubitMapping
        \EndProcedure

        \Procedure{PartitionSequencing}{P}
            \State Initialize $S_{\text{visit}} \gets \{p \mid p \in P\}$
            \State $C_{\text{components}} \leftarrow $ []

            \For{$p \in S_{\text{visit}}$} \Comment{Iterate over multiple circuits}
                
                
                %
                %
                
                \State $C_{\text{components}}.\textsc{append}(p.\textsc{Neighbours } \text{in BFS-order})$ 
                
            \EndFor
            
            \State \textbf{return} $C_{\text{components}}$
        \EndProcedure

        \Procedure{GlobalMapping}{C, $P_{\text{ordered}}$}
            \State $\pi_{C} \leftarrow $ [] \Comment{Map partitions to chiplet}
            
            \For{$C_{\text{partitions}} \in P_{\text{ordered}}$} 
                \State $p_{\text{relative}} \leftarrow \emptyset$
                
                \For{$p \in C_{\text{partitions}}$} \Comment{Iterate using the partition ordering}
                    \If{$p_{\text{relative}} = \emptyset$} \Comment{Place first partition on backend}
                        \State $\pi_{C} \leftarrow \text{C}.\textsc{PlacePartition}\text(p)$
                        \State $p_{\text{relative}} = p$
                    \Else \Comment{Place partition relative to prior}
                        \State $\pi_{C} \leftarrow \text{C}.\textsc{PlacePartitionRelative}\text(p, p_{\text{relative}})$
                    \EndIf
                \EndFor
            \EndFor
            
            \State \textbf{return} $\pi_{C}$
        \EndProcedure

        \Procedure{ChipletMapping}{C, $\pi_{C}$}
            \State $m_{v, q} \leftarrow$ [] \Comment{Virtual to physical qubit mapping}

            \For{$c \in C$} \Comment{Iterate over chiplets}
                \For{$p \in \pi_{c}$} \Comment{Iterate over partitions on chiplet}
                    \State vq $\leftarrow$ C.\textsc{MapPartition}(p) \Comment{Map qubits to partition}
                    \State $m_{v, q}.\textsc{Append}$(vq)
                \EndFor
            \EndFor

            \State \textbf{return} $m_{v, q}$
        \EndProcedure
    \end{algorithmic}
\end{algorithm}
\begin{figure*}[t]
    \centering

    \includegraphics[width=.32\linewidth]{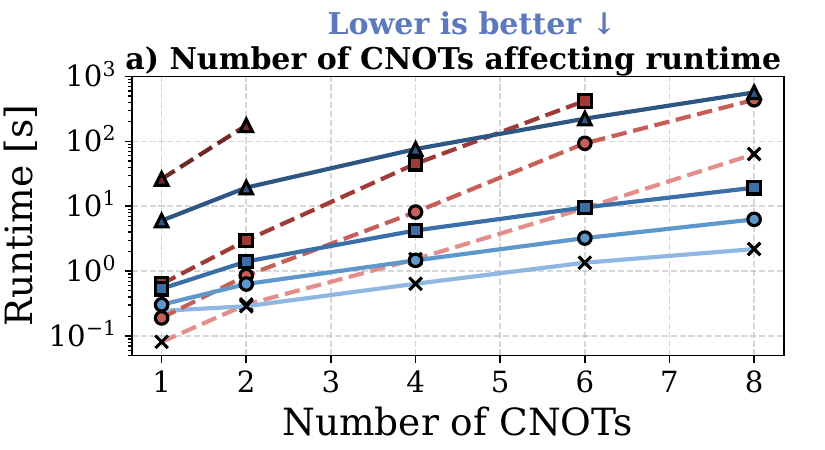}
    \includegraphics[width=.32\linewidth]{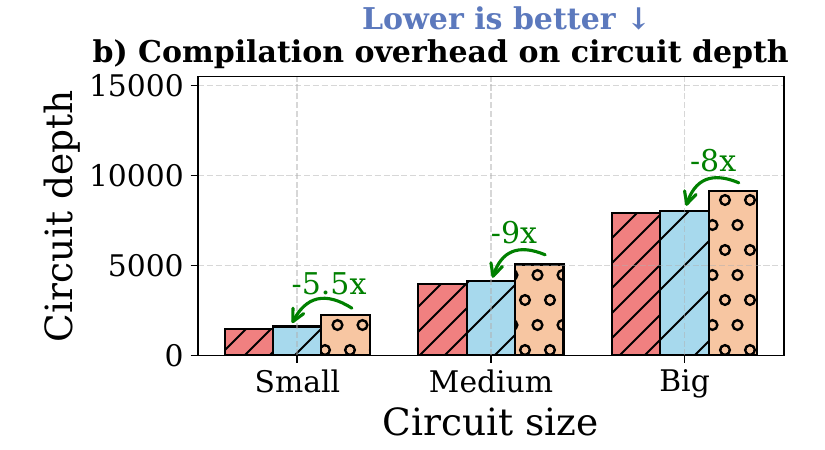}
    \includegraphics[width=.32\linewidth]{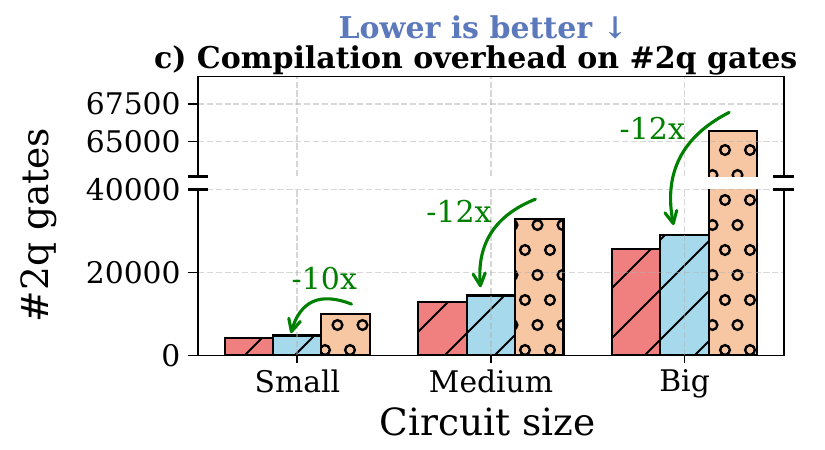}

    \vspace{0.1cm}

    \begin{minipage}[c]{0.32\linewidth}
        \hspace{-1.1cm}
        \includegraphics[width=1.85\linewidth]{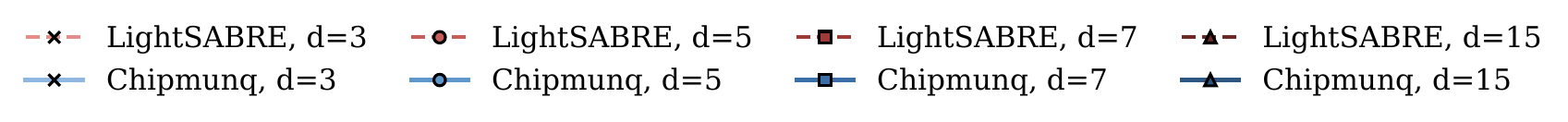}
    \end{minipage}
    \begin{minipage}[c]{0.55\linewidth}
        \hspace{4cm}
        \includegraphics[width=.6\linewidth]{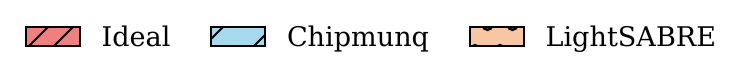}
    \end{minipage}

    \caption{Comparison between LightSABRE and \projectname{} on scaling and compilation efficiency. \emph{ \textbf{(a)} Runtime scaling for compilation of lattice surgery CNOTs to chiplet backend. Runs exceeding $10^3$s are excluded. \textbf{(b) - (c)} Circuit statistics for three circuit sizes: Small, Medium, and Big, containing 1, 3, and 6 CNOTs, respectively. \textit{Ideal} represents the initial (non-compiled) circuit.} 
    }
    \label{fig:evaluation_scalability}
\end{figure*}

        \myparagraph{(1) Sequencing}
            To assign partitions to chiplets, we require a partition ordering that specifies the order in which partitions are processed. We implement the procedure \textsc{PartitionSequencing} in Algorithm \ref{alg:implementation_mapping} by constructing the partition ordering using the \textit{BFS} (breadth-first search) algorithm applied to a partition entanglement graph. The graph is constructed from partitions as nodes and models their interactions via virtual qubit interactions.
            By visiting every node and running a BFS from each unvisited node, we define the partition order as the sequence in which nodes are discovered.
            
            In addition, we group partitions into sets, each containing partitions reachable from any other via existing dependency edges in the entanglement graph. From this disjoint set of partitions, we can construct partition orderings for multiple circuits simultaneously.

        \myparagraph{(2) Global mapping}
            Given the calculated partition order, we implement the procedure \textsc{GlobalMapping} in Algorithm \ref{alg:implementation_mapping} using a greedy bin packing problem (BP) algorithm to assign and place partitions on chiplets. Our implementation follows the general heuristic approach commonly applied in related work \cite{ccmap, route_forcing_escofet}: Assign as many partitions as possible to each chiplet to minimize inter-chiplet communication between partitions on different chiplets.
            
            We reformulate the partition assignment problem as a 2-dimensional BP (2BP) \cite{2d_bpp}, where chiplets act as bins containing rectangular regions. We implement two variants of the first-fit algorithm \cite{first_fit_bpp} for solving the 2BP, differing in the placement of the first partition on a chip bin: \textit{center} places it in the center of the selected chiplet, while \textit{size-aware} employs a top-left approach based on the bottom-left heuristic \cite{2d_bpp}, where the partition is placed at the top-left as far as possible. We propose \textit{center} for a one-to-one allocation with partition sizes corresponding to chiplet sizes, whereas \textit{size-aware} is suitable for a many-to-one strategy where multiple partitions are assigned to a single chiplet. We implement this approach in the \textsc{PlacePartition} procedure in Algorithm \ref{alg:implementation_mapping} which has complexity $\mathcal{O}(N^2)$ relative to the number of partitions $N$ \cite{2d_bpp}
            
            After assigning a partition to a chiplet bin, we remove the corresponding region and perform a recursive guillotine cut \cite{guillotine_bpp} to recalculate free regions. An example of the process is shown in Fig. \ref{fig:chiplet_mapping}.
            
            For partitions that interact with each other, we implement a tight placement approach \textsc{PlacePartitionRelative} in Algorithm \ref{alg:implementation_mapping}, resulting in reduced routing distances. Our implementation selects the nearest available region on the same chip bin as the reference partition. If no free region is available, we pick the closest chip bin that can fit the current partition. We extract the relative orientation of partitions from the labels of the virtual qubits in the logical circuit. By comparing the qubit labels between relative partitions, we can determine whether a partition should be placed below or to the right of the relative partition.
            
            To ensure that the mapping phase is aware of defective qubits, we extend the chiplet bins with \textit{no-placement zones}. These zones represent the locations of the defective qubits that cannot be used and are accounted for in the 2BP problem. 
            
        \myparagraph{(3) Local mapping}
            Lastly, it is necessary to translate the 2BP assignment into physical qubits.
            The 2BP assignment maps partitions to chiplets, with local chiplet coordinates indicating the location of each partition on each chiplet. We implement the chiplet mapping in the \textsc{LocalMapping} procedure in Algorithm \ref{alg:implementation_mapping}. 
            By using both the partition size and local coordinates, we determine the corresponding physical qubits assigned to this partition. We iterate over all chiplets and partitions and construct a mapping between virtual and physical qubits.

    \subsection{Routing}\label{subsec:implementation_routing}
        Following the mapping of virtual to physical qubits, we need to ensure that, whenever two qubits interact in the initial circuit, they are placed in physical qubits that can interact with each other. We fix the calculated qubit mapping and construct a new quantum circuit with additional SWAP operations between distant physical qubits to satisfy connectivity constraints.

        To maximize efficiency, we employ both \textit{intra-chiplet} and \textit{inter-chiplet} routing strategies. For \textbf{intra-chiplet} routing, we implement a greedy shortest-path routing procedure for qubits on the same chiplet.
        Given the construction of FT circuits, patches are designed to avoid any additional routing. Thus, we ignore qubits of the same partition.
        For partition interactions, we employ brute-force $\text{SWAP}$ routing along the shortest path. This path is bifurcated, and routing is simultaneously performed from both qubits to the midpoint. We calculate the shortest unweighted path using Dijkstra's algorithm \cite{dijkstra}. While more sophisticated routing techniques are available, we utilize this method because our mapping approach yields short path lengths that do not necessitate additional complexity.
        
        To enable \textbf{inter-chiplet} routing, we implement a chiplet-aware routing procedure that accounts for heterogeneous noise levels of inter-chiplet links. We formulate a total cost function consisting of three interdependent components derived from the most important hardware-aware metrics:

        \begin{equation}
            \small
            \begin{aligned}
                \textsc{PathCost}(P,\hspace{.1cm} q_{icc}) = 
                \underbrace{|P|}_{\text{Path length}} 
                 + \hspace{.2cm} \underbrace{\alpha \cdot ICC(q_{icc})}_{\text{Inter-chiplet cost}} 
                 \hspace{.2cm} + \underbrace{\beta \cdot U(q_{icc})}_{\text{Congestion penalty}}
            \end{aligned}
            \label{eq:global_routing_path_cost}
        \end{equation}

        consisting of the length of the shortest path $P$ between source and target qubits utilizing the $q_{icc}$ inter-chiplet link, $ICC(q_{icc})$ as the noise level of the inter-chiplet link, and $U(q_{icc})$ the number of utilizations of the inter-chiplet link. We use $\alpha$ and $\beta$ as tunable hyperparameters: while $\alpha$ prioritizes selecting the inter-chiplet connection with the highest fidelity, $\beta$ prioritizes selecting the path that uses the least-congested inter-chiplet connections.

        Given an initial \textit{shortest} path between qubits of different chiplets, we select the inter-chiplet connection resulting in the lowest path cost according to Eq.~\ref{eq:global_routing_path_cost}. We identify the $k$-nearest inter-chiplet connections adjacent to the current selection. After calculating the total cost for each, the path with the minimum cost is selected, and the corresponding inter-chiplet connection utilization counter is incremented to reflect its usage.
        
        By varying the parameters $\alpha$ and $\beta$ in Eq.~$\text{\ref{eq:global_routing_path_cost}}$, we enable fine-grained control over the routing focus, balancing competing metrics such as code distance reduction and increased circuit depth.    
        The optimal values of $\alpha$ and $\beta$ depend heavily on the specific circuit structure and require fine-tuning based on the inter-chiplet noise.

        Our routing stage exhibits a worst-case runtime complexity of $\mathcal{O}((n_{gl} + k \cdot n_{gr}) E \log V)$ accounting for local $n_{gl}$ and remote $n_{gr}$ gates, with $E$ total number of edges across all chiplets, $V$ total number of qubits across the entire system, and $k$ as the number of inter-chiplet connections considered during inter-chiplet routing \cite{dijkstra}.

\begin{figure*}[t]
    \centering

    \begin{minipage}[c]{1\linewidth}
        \centering
        \includegraphics[width=.32\linewidth]{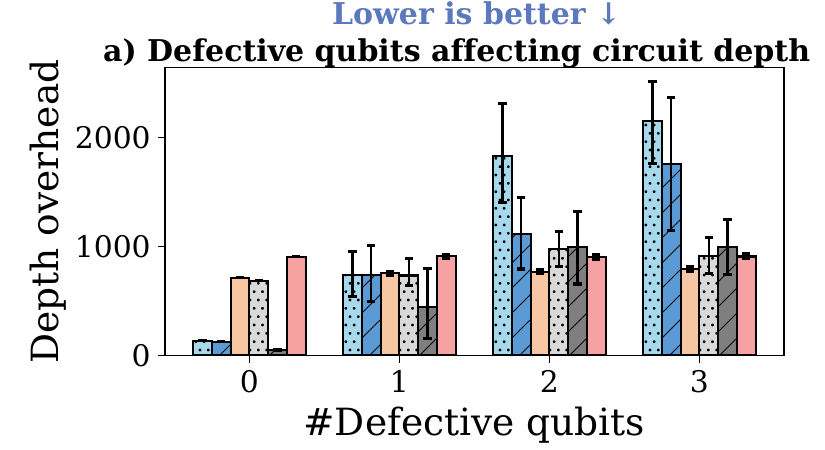}
        \includegraphics[width=.32\linewidth]{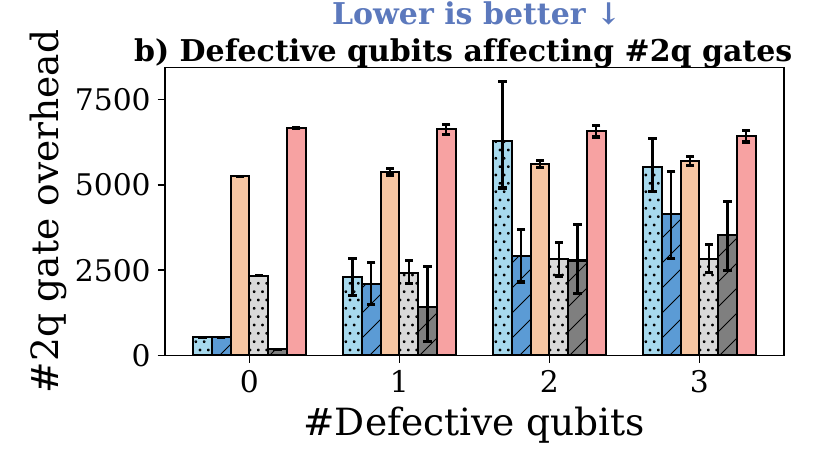}
        \includegraphics[width=.32\linewidth]{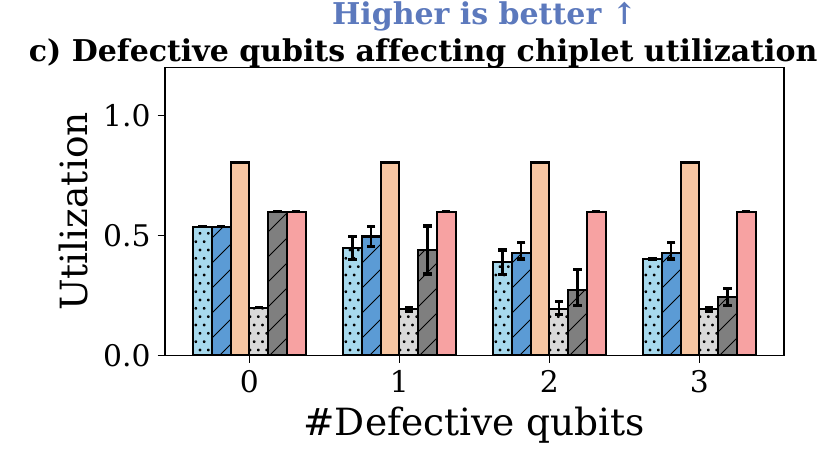}
    \end{minipage}
    

    \begin{minipage}[c]{1\linewidth}
        \centering
        \includegraphics[width=.85\linewidth]{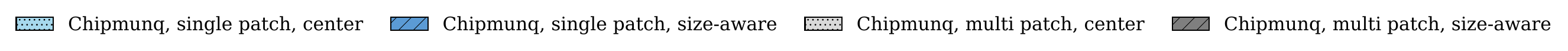}
        
        \vspace{-.15cm}
        \includegraphics[width=.35\linewidth]{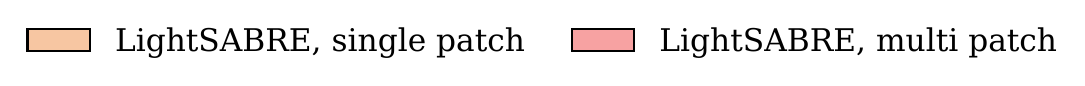}
    \end{minipage}

    \vspace{-4pt}
    \caption{Performance impact of different compilation approaches for varying numbers of defective qubits per chiplet. \emph{ Effect of compilation approach on \textbf{(a)} circuit depth overhead, \textbf{(b)} two-qubit gate overhead, and \textbf{(c)} chiplet utilization. 
    }}
    \label{fig:evaluation_defective_qubits}
    \vspace{-5pt}
\end{figure*}
    \section{Implementation and Evaluation}
    We evaluate our compiler across four primary dimensions: \textbf{(1)} We analyze performance through runtime scalability and circuit statistics (\S~\ref{subsec:evaluation_performance}), \textbf{(2)} we examine the architectural impact of inter-chiplet connectivity and hardware defects (\S~\ref{subsec:evaluation_chiplet}), \textbf{(3)} we assess QEC fidelity by simulating LER variations across diverse backend configurations (\S~\ref{subsec:evaluation_qec}), and \textbf{(4)} investigate the specific influence of routing on the LER to understand the trade-offs imposed by chiplet-constrained interconnects (\S~\ref{subsec:evaluation_routing}).

    \subsection{\projectname{} Implementation}
        We integrate \projectname{} into Qiskit v2.30rc1 \cite{qiskit} in Python v3.12 by constructing a custom pass manager pipeline using the {\scshape \textsc{qiskit.StagedPassManager}} instance consisting of three {\scshape \textsc{qiskit.PassManager}} instances implementing the partitioning, mapping, and routing stages. We use networkX v2.8.8 \cite{networkx} in combination with the KaHyPar v1.3.6 \cite{kahypar} for circuit partitioning. Our global routing implementation utilizes Dijkstra's algorithm implemented in rustworkx v0.17.1 \cite{rustworkx}.

    \subsection{Experimental Methodology}
        \myparagraph{Experimental setup}
            We use Stim v1.15.0 \cite{stim_gidney} and PyMatching v2.3.1 \cite{mwpm_pymatching} for performing circuit simulations, with $10^8$ shots per run, and utilize qiskit-qec \cite{qiskit_qec} for translating qiskit circuits to the Stim format \cite{stim_gidney}. All experiments are performed on a Linux machine running NixOS 25.11, equipped with one AMD EPYC 7713P 64-Core Processor and 1 TiB of main memory.

        \myparagraph{Compilation input}
            We use tQEC v0.1.0 \cite{tqec} to generate surface-code-protected lattice surgery circuits with circuit-level noise based on the SI1000 error model \cite{si1000_error_model}, accounting for both inter- and intra-chip gates. Large-scale FT circuits are represented by lattice-surgery CNOT operations, achieved by duplicating the initial circuit and adjusting the code distance $d$. Since logical code patches in FT circuits are inherently defined during circuit construction, we bypass the partitioning stage and utilize circuits with pre-defined partitions, reflecting anticipated inputs for future FT compilers. In all experiments, backends contain more chiplets than patches to avoid space-constrained placement, with each chiplet dimensioned at 30\% above the resources required by a single code patch, enabling one-to-one assignment flexibility.
        
        \myparagraph{Metrics}
            We evaluate \projectname{} using the following metrics: \textbf{(1) Gate overhead:} The ratio between the total number of gates in the compiled circuit and the number of gates in the original logical circuit. \textbf{(2) Depth overhead:} The ratio between the critical path length of the compiled circuit and that of the original circuit. \textbf{(3) LER:} The simulated probability that the compiled circuit produces an incorrect logical outcome, accounting for both physical gate errors and the effects of QEC.
        
        \myparagraph{Baselines}
            We select LightSABRE \cite{ibm_sabre}, part of Qiskit \cite{qiskit}, as our baseline due to superior scalability for large FT circuits (\S~\ref{sec:motivation}) and to enable comparability with literature \cite{original_sabre, swin, route_forcing_escofet, qecc_synth}. We exclude QECC-Synth \cite{qecc_synth} due to scalability constraints and omit MECH \cite{mech} due to excessive gate overhead and format incompatibility (\S~\ref{sec:motivation}).

\begin{figure*}[t]
    \centering

    \begin{minipage}[c]{0.32\linewidth}
        \centering
        \includegraphics[width=\linewidth]{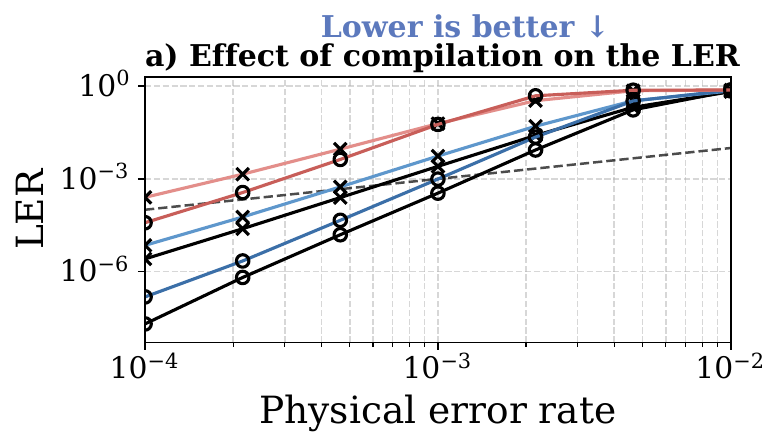}
    \end{minipage}
    \hfill
    \begin{minipage}[c]{0.67\linewidth}
        \centering
        \includegraphics[width=0.49\linewidth]{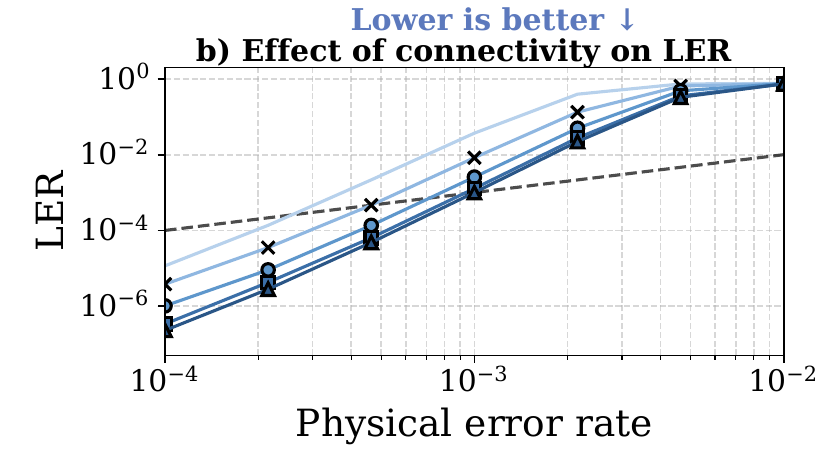}
        \includegraphics[width=0.49\linewidth]{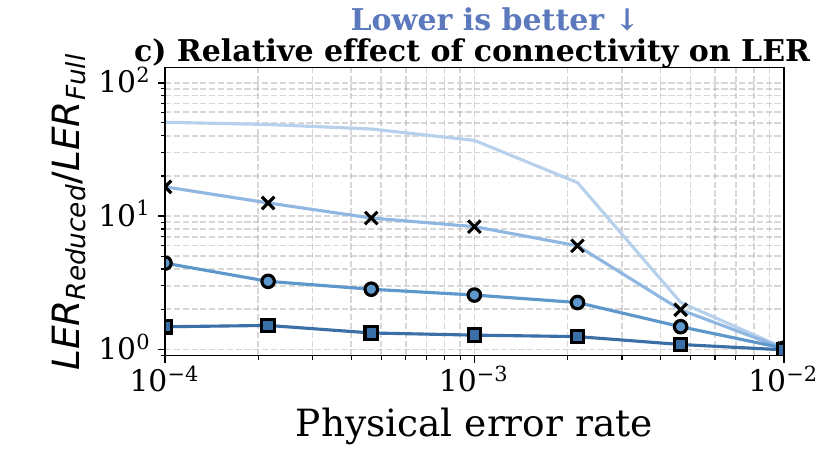} 
    \end{minipage}

    \vspace{0.1cm}

    \begin{minipage}[c]{0.32\linewidth}
        \hspace{-1.2cm}
        \includegraphics[width=1.3\linewidth]{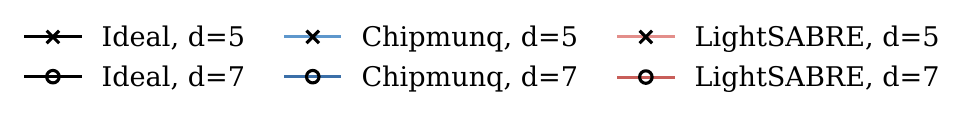}
    \end{minipage}
    \begin{minipage}[c]{0.55\linewidth}
        \centering
        \vspace{-.1cm}
        \hspace{1cm}
        \includegraphics[width=.59\linewidth]{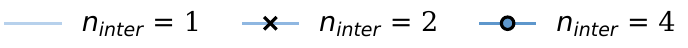}

        \vspace{-.1cm}
        \hspace{1cm}
        \includegraphics[width=.4\linewidth]{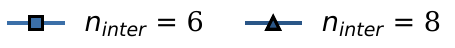}
    \end{minipage}

    \vspace{-5pt}
    \caption{Effect of patch distribution on logical error rate of the CNOT lattice surgery operation for $p_{inter} = 10^{-3}$. \emph{ \textbf{(a)} Effect on the LER after compilation using \projectname{} and LightSABRE for $d \in \{5, 7\}$. \textbf{(b)} LER after compilation of the circuit with $d=7$ to a backend with a varying number of inter-chiplet links. \textbf{(c)} Increase of the LER for connectivity-constrained backends over the $n_{inter} = 8$ run. }}
    \label{fig:evaluation_qec_inter_chiplet}
    \vspace{-5pt}
\end{figure*}

    \subsection{\projectname's End-to-End Performance} \label{subsec:evaluation_performance}
        
        \textbf{RQ\#1 (Scalability):} {\em How does the runtime of the proposed implementation scale as circuit complexity increases?}

        Practical FT operations are projected to require code distances of $d \approx 27$ \cite{Acharya2025}. At this scale, the resource requirements result in significant computational overhead during compilation. Consequently, compilers need to demonstrate tractable runtime scaling.
        
        \myparagraph{Methodology} We evaluate \projectname{} by compiling circuits composed of CNOT lattice-surgery instructions, each comprising three surface code patches, to chiplet backends. We assess performance by scaling both the instruction count and the code distance. 

        \myparagraph{Results} 
        As shown in Fig.~\ref{fig:evaluation_scalability} (a), our implementation significantly outperforms LightSABRE, achieving an average speedup of 13.5$\times$ (up to 70$\times$ for $d=7$ and 8 CNOTs), and compiles 4$\times$ more CNOTs for $d=15$. Chipmunq exhibits a flatter runtime curve, resulting in greater speedups as circuit size and complexity increase. For \projectname{}, mapping scales with qubit count only, whereas routing scales with both qubit count and gate density, resulting in increased routing duration as circuit complexity increases.
                    
        \takeaway{\myparagraph{Takeaway \#1} \projectname{} reduces the compilation runtime by $13.5\times$ on average compared to LightSABRE, providing increased scalability for large-scale circuits with high code distances.}

        \textbf{RQ\#2 (Circuit complexity):} {\em How do circuit depth and gate overhead scale as circuit size increases?}
        
        Compilers that ignore QEC patch structures incur routing overhead, introducing extra noise that reduces code effectiveness. Thus, minimizing circuit depth and gate overhead is critical.
               
        \myparagraph{Methodology} We generate circuits of increasing size by scaling the number of patches and evaluate the resulting compiled circuits in terms of depth overhead and gate overhead relative to their ideal (unmapped) counterparts. Circuit structure and backend configuration are identical to those used for RQ\#1 to enable direct comparison.

        \myparagraph{Results} Fig.~\ref{fig:evaluation_scalability} (b) and (c) show that \projectname{} achieves an average $7.5 \times$ and $11.3 \times$ reduction in circuit depth and gate count across multiple circuit sizes, respectively, compared to LightSABRE.

        \takeaway{\myparagraph{Takeaway \#2} \projectname{} cuts circuit depth by $7.5 \times$ and gate overhead by $11.3 \times$ compared to LightSABRE, enabling stronger error protection.}

    \subsection{Chiplet-suitability}\label{subsec:evaluation_chiplet}
       
        \textbf{RQ\#3 (Defects):} {\em How do defective qubits affect performance?}

        Chiplet fabrication is subject to inherent variability, where the probability of defects scales with the total chip count \cite{monolithic_to_chiplets}. To maximize chiplet utilization, compilers must be defect-aware to make the most of the remaining resources.
        
        \myparagraph{Methodology} We use the CNOT circuit from RQ\#1, consisting of three patches, and compile it for a backend with a varying number of defective qubits. To evaluate the effect of defective qubits using the methods \textit{center} and \textit{size-aware}, we consider two chiplet sizes: \textit{single patch}, where one chiplet can fit one patch (plus some additional spacing), and \textit{multi patch}, where one chiplet can fit four patches. We calculate the chiplet utilization as $N_{\mathrm{qubits}} / (N_{\mathrm{qubits\ per\ chiplet}} \cdot N_{\mathrm{chiplets}}) \in (0, 1]$, where a maximum utilization of 75\% is achievable under our circuit and backend configuration.
 
        \myparagraph{Results} As shown in Fig.~\ref{fig:evaluation_defective_qubits} (a) and (b), increasing defective qubits strongly influences depth and two-qubit gate overhead during compilation. LightSABRE consistently produces circuits with up to 2$\times$ more overhead compared to \projectname{}, but achieves near-maximal chiplet utilization by routing individual qubits around defective regions rather than placing entire contiguous patches.
        For \projectname{}, in the \textit{single patch} configuration, chiplets with more than one defective qubit result in up to 50\% increase in both depth and gate overhead compared to no defects. However, \textit{size-aware} placement reduces this overhead by up to 40\% compared to \textit{center} placement. The \textit{multi patch} configuration exhibits comparable overhead across all placement methods. As illustrated in Fig.~\ref{fig:evaluation_defective_qubits} (c), defective qubits can reduce chiplet utilization by up to 50\% compared to no defects. The \textit{size-aware} method improves average utilization by 9\% across both configurations compared to \textit{center} placement.
            
        \takeaway{\myparagraph{Takeaway \#3} \projectname{} reduces circuit overhead by 2$\times$ and backend utilization by 20\% on average compared to LightSABRE, thus enabling high-quality compilation on defective backends.}
        
\begin{figure*}[t]
    \centering
    \includegraphics[width=.32\linewidth]{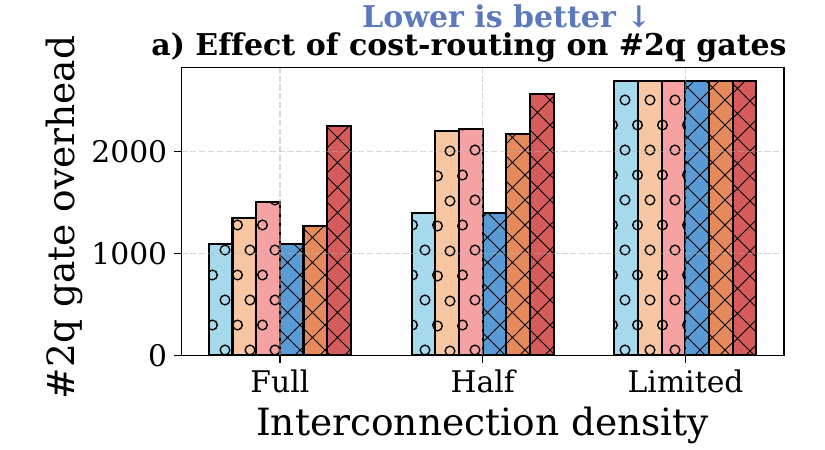}
    \includegraphics[width=.32\linewidth]{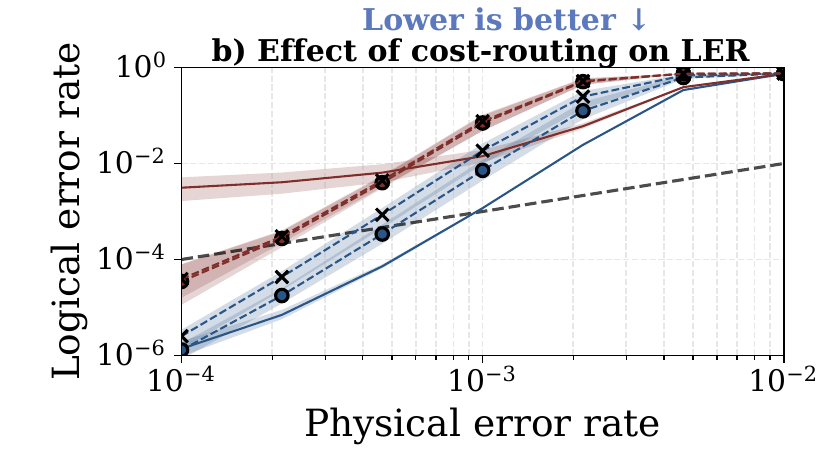}
    \includegraphics[width=.32\linewidth]{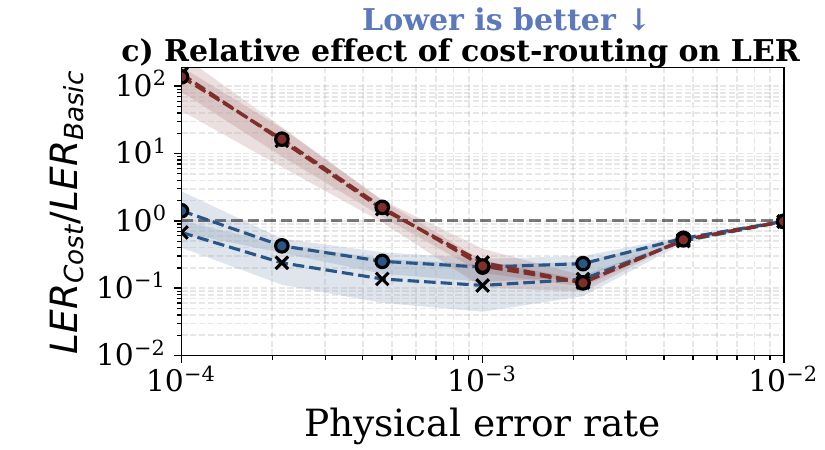}

    \begin{minipage}[c]{0.32\linewidth}
        \hspace{-1.1cm}
        \includegraphics[width=1.4\linewidth]{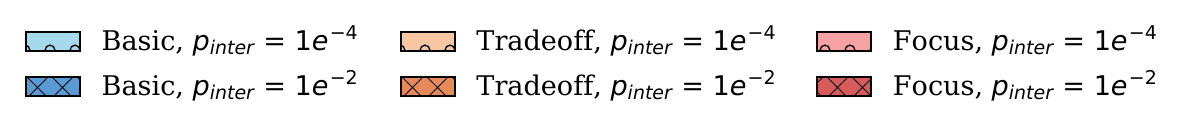}
    \end{minipage}
    \begin{minipage}[c]{0.55\linewidth}
        \hspace{1.5cm}
        \includegraphics[width=.95\linewidth]{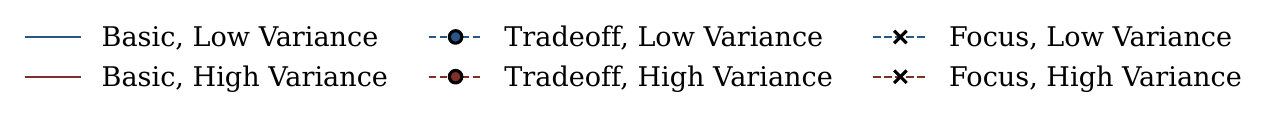}
    \end{minipage}
    \centering

    \vspace{-4pt}
    \caption{Effect of cost routing on the circuit properties and LER on Surface code with $d=7$.
        \emph{
            \textbf{(a)}
            Two-qubit gate overhead on a chiplet backend with randomized link errors across and variable inter-chiplet connections, using both high and low noise levels.
            \textbf{(b)} Effect of three routing policies on LER for a high and low variance inter-chiplet link noise model under a medium noise level.
            \textbf{(c)} Change in LER by using either the focus or tradeoff over the basic configuration.
        }  
    }
    \vspace{-5pt}
    \label{fig:evaluation_routing}
\end{figure*}

    \subsection{QEC-suitability} \label{subsec:evaluation_qec}
        
        \textbf{RQ\#4 (Distribution \& Connectivity):} {\em How does distributing patches across chiplets with limited inter-chiplet connectivity affect logical error rates?}
        
        \myparagraph{Methodology} We compile CNOT circuits from RQ\#1 with $d \in \{5, 7\}$ to backends where patches are distributed across multiple chiplets with fully-connected inter-chiplet links ($p_{inter} = 10^{-3}$). We then evaluate $d=7$ circuits across backends with varying inter-chiplet link counts: $n_{inter} \in \{8, 6, 4, 2, 1\}$. We measure LER for the \textit{ideal} (non-compiled) circuit, \projectname{}, and LightSABRE.

        \myparagraph{Results} Fig.~\ref{fig:evaluation_qec_inter_chiplet} (a) shows that LightSABRE increases LER by $128\times$ on average compared to the \textit{ideal} circuit, failing to maintain error suppression. \projectname{} reduces LER by $56\times$
        compared to LightSABRE while increasing LER by only $2.2\times$ on average relative to \textit{ideal}. Higher code distances amplify compilation impact ($2\times$ at $d=3$, up to $7\times$ at $d=7$), though exponential error suppression renders this negligible. When inter-chiplet connectivity is constrained (Fig.~\ref{fig:evaluation_qec_inter_chiplet} (b, c)), \projectname{} maintains low LER ($1.5\times$ increase) when $n_{inter}>d$. Below this threshold ($n_{inter} < d$), LER increases by up to $50\times$ for $n_{inter}=1$, indicating code distance acts as a connectivity bottleneck.
        
        \takeaway{\myparagraph{Takeaway \#4} \projectname{} reduces LER by $160\times$
        compared to LightSABRE and maintains $3\times$ average overhead versus \textit{ideal} circuits. For connectivity-constrained chiplets, \projectname{} preserves error suppression when $n_{inter} \geq d$, with $18\times$ average LER reduction even below this threshold.}

    \subsection{Routing for QEC} \label{subsec:evaluation_routing}
        \textbf{RQ\#5 (Connectivity \& routing):} {\em How do inter-chiplet connectivity constraints and noise-aware routing strategies influence circuit overhead and logical error rates?}

        \myparagraph{Methodology} We use the circuit from RQ\#1 and evaluate three interconnection densities: $n_{inter} \in \{8, 4, 1\}$ (\textit{full}, \textit{half}, \textit{limited}), where \textit{full} connects all adjacent edge qubits, \textit{half} reduces density by 50\%, and \textit{limited} uses a single link between neighboring chiplets. We define two inter-chiplet noise levels — high ($p_{inter} = 10^{-2}$) and low ($p_{inter} = 10^{-4}$) — randomly scaled in $[1, 10]$. For LER evaluation ($d=7$, fully-connected chiplets), we test two variance configurations: \textit{low variance} $[1,10]$ and \textit{high variance} $[1,100]$. We evaluate three routing methods: \textit{Basic} (noise-agnostic), \textit{Focus} (selects lowest-error links), and \textit{Tradeoff} (balances noise and path length).

        \myparagraph{Results} Fig.~\ref{fig:evaluation_routing}~(a) shows that reducing connectivity increases circuit depth and two-qubit gates by 10\% and 23\%, respectively, for \textit{Basic} routing. Under \textit{limited} configuration, gate overhead increases $2.6\times$ versus \textit{full}, while \textit{full} and \textit{half} exhibit marginal overhead. \textit{Tradeoff} and \textit{Focus} incur 30\% and 40\% additional overhead over \textit{Basic}, reflecting link prioritization. Consistent with RQ\#5, overhead remains negligible when $n_{inter} > d$. High inter-chiplet noise increases gate overhead by 50\% as the algorithm prioritizes higher-fidelity links, leading to longer paths. For LER (Fig.~\ref{fig:evaluation_routing} (b, c)), \textit{Focus} and \textit{Tradeoff} reduce LER by up to $11\times$ on average versus \textit{Basic} in high-variance environments, despite increased overhead. Low-variance and high physical-error-rate scenarios favor \textit{Basic}'s full link utilization. Hyperparameter sweeps confirm consistent performance gains.

        \takeaway{\myparagraph{Takeaway \#5} Inter-chiplet link reduction is constrained by code distance $d$: negligible overhead when $n_{inter} > d$. In high-variance noise environments, \projectname{}'s \textit{cost-routing} achieves $10\times$ LER reduction versus noise-unaware routing, enabling reliable execution on heterogeneous chiplet hardware.}

    \section{Related work}
    The quantum circuit mapping and routing problem is a widely studied area of research, with most prior work focusing on monolithic architectures for general-purpose quantum circuits and focusing on reducing the number of SWAP operations and circuit depth. Recent work incorporates additional constraints by addressing different hardware topologies and by considering special classes of quantum circuits, such as FT circuits.

    \myparagraph{General mapping and routing} Among these algorithms, SABRE and its optimized successor, LightSABRE, are widely recognized as the first scalable algorithms for the qubit mapping problem \cite{original_sabre, ibm_sabre}. Fu et al. extend this line of approach by introducing a noise-aware variant that incorporates hardware noise, yielding improved performance across various benchmarks \cite{swin}.
        
    \myparagraph{Mapping of QEC codes} Yin et al. propose QECC-Synth, an automated compiler that maps circuits defined using stabilizer codes to a variety of hardware topologies \cite{qecc_synth}. Their implementation is limited to hundreds of qubits due to the utilization of a SAT solver.
    Watkins et al. propose liblsqecc, a large-scale compiler for performing lattice surgery \cite{Watkins_2024}. While the implementation does not map to hardware and thus cannot be executed, it offers the scalability necessary for large-scale FT circuits.
    tQEC is a design automation software library for generating lattice-surgery instructions for surface code, but does not schedule surgery operations \cite{tqec}.

    \myparagraph{Mapping and routing to chiplet architecture} Zhang et al. propose MECH, a routing algorithm that uses ancillary qubits to form a computational highway across chiplets, combining both gate-based and measurement-based computing  \cite{mech}.
    In contrast, Bandic et al. reformulate the problem as a k-partitioning problem and solve this using quadratic unconstrained binary optimization \cite{mapping_qubo}, while assuming full qubit connectivity within and across QPUs.
    Du et al. incorporate hardware constraints such as the noise level of inter-chiplet connections and leverage them in a hardware-aware routing implementation for general-purpose circuits \cite{ccmap}. 
    In addition, Escofet et al. propose an approach that combines routing and mapping using a force-directed formulation, accounting for the attraction forces between interacting qubits \cite{route_forcing_escofet}.

    \takeaway{\textbf{How our paper differs?}
        To the best of our knowledge, we provide the first hardware-aware compiler for mapping and routing fault-tolerant circuits onto chiplet-based superconducting architectures, applicable to a wide range of topological QEC codes.
    }

    \section{Conclusion}
    \projectname{} successfully bridges the compilation gap between fault-tolerant quantum circuits and modular chiplet architectures by treating QEC patches as first-class citizens throughout the mapping and routing process. By explicitly accounting for inter-chiplet link heterogeneity and patch geometry, \projectname{} scales quantum circuits without the typical explosion in gate overhead or logical errors that plague qubit-level compilation approaches. Our evaluation demonstrates that modularity does not have to compromise fidelity: patch-aware partitioning, staged dependency-aware placement, and cost-aware routing collectively enable both scalability and error suppression. 

    \myparagraph{Artifact} We release an open-source implementation of \projectname{} on Github at \url{https://github.com/Wegii/chipmunq}.

    \section*{Acknowledgment}
        This work was funded by the Bavarian State Ministry of Science and the Arts as part of the Munich Quantum Valley (MQV) initiative, grant number 6090181.
    
    \bibliographystyle{IEEEtran}
    \bibliography{bibliography}

\end{document}